\documentclass[12pt]{article}

\usepackage{graphicx}

\def \lsim{\mathrel{\vcenter
     {\hbox{$<$}\nointerlineskip\hbox{$\sim$}}}}
\def \gsim{\mathrel{\vcenter
     {\hbox{$>$}\nointerlineskip\hbox{$\sim$}}}}

\newcommand{\bH}{{\bf H}}
\newcommand{\bh}{{\bf h}}
\newcommand{\beq}{\begin{equation}}
\newcommand{\eeq}{\end{equation}}
\newcommand{\beqa}{\begin{eqnarray}}
\newcommand{\eeqa}{\end{eqnarray}}
\newcommand{\beqar}{\begin{eqnarray*}}
\newcommand{\eeqar}{\end{eqnarray*}}

\begin{document}
\thispagestyle{empty}

\hfill{\sc UG-FT-189/05}

\vspace*{-2mm}
\hfill{\sc CAFPE-59/05}

\vspace{32pt}
\begin{center}

\textbf{\Large 
Higgsino dark matter in partly supersymmetric models}
\vspace{40pt}

M. Masip and I. Mastromatteo
\vspace{12pt}

\textit{
CAFPE and Departamento de F{\'\i}sica Te\'orica y del
Cosmos}\\ \textit{Universidad de Granada, E-18071, Granada, Spain}\\
\vspace{16pt}
\texttt{masip@ugr.es, iacopomas@infis.univ.trieste.it}
\end{center}

\vspace{40pt}

\date{\today}

\begin{abstract}

Models where supersymmetry (SUSY) is manifest only in a sector
of the low-energy spectrum have been recently proposed as an
alternative to the MSSM. In these models the electroweak scale 
is explained by 
a fine-tuning between different Higgs mass contributions 
({\it split-SUSY models}), or 
by the localization of the Higgs sector in a point 
of an extra dimension where all the mass parameters are 
suppressed by the metric 
({\it partly-SUSY models}). Therefore, the presence of a good dark
matter candidate becomes the main motivation for (partial)
low-energy SUSY. 
We study this issue in minimal frameworks where the higgsinos
are the only light supersymmetric particles. 
Whereas in split-SUSY models the higgsino should have a mass 
around $1$ TeV, we show that in partly-SUSY models the 
lightest higgsino could also be found below $M_W$.

\end{abstract}


\newpage

\section{Introduction}

The stability of the electroweak (EW) scale at the loop level
has been the main motivation for supersymmetry (SUSY) 
during the past 25 years \cite{Sohnius:1985qm}. 
SUSY {\it doubles} the spectrum of the standard 
model (SM) and makes the EW scale {\it natural}, consistent 
with the dynamics and not the result of an
accidental cancellation between higher scales.

Recently, however, other alternatives have been proposed where
SUSY does not play this traditional role. 
We will consider two frameworks:

{\it (i)} Split-SUSY models \cite{Arkani-Hamed:2004fb}, 
where the Higgs mass parameters are 
fine-tuned to the actual values, which provide an EW scale that 
allows atoms \cite{Agrawal:1997gf}. An analogous argument was 
used by Weinberg \cite{Weinberg:1987dv}
to predict a cosmological constant consistent with the observed
value (that allows the formation of structures), 
and it would be justified by the 
{\it landscape} \cite{Bousso:2000xa} of string theory.

{\it (ii)} Partly-SUSY models \cite{Gherghetta:2003wm}, 
with the Higgs sector living in a
point of an extra dimension (the {\it TeV brane}) where the 
Planck scale is red-shifted by the metric to the EW scale.
SUSY is then broken in the {\it Planck brane}.
Loop corrections connect the Higgs with Planck-scale physics,
but all the contributions are also suppressed by the metric. 
These models have a 4-dimensional (4D) holographic interpretation 
where the Higgs is a bound state of size TeV$^{-1}$ and its
constituents decouple exponentially at energies above that 
scale.

Although in these two
frameworks SUSY would be part of the complete theory
in the ultraviolet, it is not needed 
to cancel large quadratic corrections 
and could be broken at very high energies.
In both cases, however, the breaking may
be such that one is left at low energies with what has been
an important phenomenological motivation for SUSY: the 
presence of a good dark matter candidate \cite{Goldberg:1983nd}. 
A stable, weakly-interacting particle like the neutralinos 
of the MSSM could provide a relic abundance  
$\Omega_{\chi}=0.113 h^2$ \cite{Bennett:2003bz}, in agreement 
with cosmological and astrophysical observations.

One can argue that, in both scenarios, the higgsino 
would be a well motivated lightest SUSY particle (LSP).  
The additional presence of gauginos at a low scale 
requires that the breaking of SUSY respects 
an (approximate) R-symmetry. In \cite{Arkani-Hamed:2004yi}
it is shown that this can be naturally the case when 
SUSY is broken by the nonzero $D$ term of a spurion 
vector superfield. However, in the generic case with $D$- 
and $F$-breaking the gauginos should get large masses.
In contrast, in partly-SUSY models SUSY may be broken
at a very large scale in the Planck brane, but the $\mu$ term 
(localized in the TeV brane) will always have TeV size.
In split-SUSY models a $\mu$ term of order TeV could
be obtained, for example, in a gauge-mediated scenario
\cite{Giudice:1998bp} where SUSY is broken at a scale 
$F\approx (10^{11}$ GeV$)^2$. For messenger masses of 
the same order, scalar and gauginos will get large masses  
$\tilde m\approx (\alpha/4\pi) \sqrt{F}$ 
through gauge interactions, whereas the $\mu$ term 
would be of order TeV if it is induced just by gravitational 
interactions \cite{Giudice:1988yz}.
It could also be that SUSY is broken at 
the grand unification (GUT) scale but $\mu$ is 
still protected by flavour symmetries of the superpotential
and the K\"ahler potential. 
These symmetries 
(for example, the discrete symmetries related to the topology 
of the compact space in string models \cite{Greene:1986bm}) 
are suggested by the hierarchies in the quark and lepton 
masses. Notice that at large scales one may expect several
pairs of higgs doublets (per just one gaugino multiplet), 
so the symmetries could protect (at least) one of them 
and imply a $\mu$ term of order TeV generated by 
non-renormalizable operators \cite{Kim:1983dt}.

Here we explore this {\it minimal} possibility, with 
the higgsinos as the only light SUSY particles.
We focus our analysis on higher order effects (loops and
heavy fields) that may break the degeneracy of the four
higgsinos. We show that although split- and partly-SUSY 
models may look similar, they suggest a very different 
higgsino dark matter scenario. 

The dark matter candidates in generic 
split-SUSY models have been analized in \cite{Pierce:2004mk},
with results that do not differ essentially from the ones
in the MSSM. Studies of indirect dark matter detection signals
can be found in \cite{Arvanitaki:2004df}.
Arkani-Hamed {\it et al.} \cite{Arkani-Hamed:2004yi}
have suggested an interesting possibility where a heavy gravitino
decays when the LSP is already out of equilibrium, increasing 
its cosmic abundance. Here we will just determine the usual 
relic abundance obtained from a LSP with (approximately) 
constant number below the freeze-out temperature.

\section{Higgsinos in the split-SUSY model}

Let us start defining the spectrum in the split-SUSY model.
We will asume that sfermions and gauginos get masses at the 
SUSY-breaking scale, much above the 
EW scale. The Higgs sector contains the usual 
doublets of chiral superfields, 
$\bH_1=(\bh_1^0\; \bh_1^-)$ and 
$\bH_2=(\bh_2^+\; \bh_2^0)$, that can be expanded 
\beq
\bh_1^0= h_1^0 + \sqrt{2}\; \theta \tilde h_1^0 + \theta^2 F_{h_1^0}\;,
\label{eq1}
\eeq
with $h_1^0$, $\tilde h_1^0$ and $F_{h_1^0}$ the scalar, spinor 
and auxiliary components of $\bh_1^0$
and analogous expressions for the rest of higgs fields. 
Once the SUSY-breaking terms are included, the light  
scalar sector will coincide with the one in the SM (all the 
extra scalars become very heavy). In the higgsino sector,
we assume a term $W=\mu \bH_1 \bH_2$ in the superpotential,
giving 
\beqa
{\cal L}_0 &=&\int {\rm d}^2\theta\; \mu\; (-\bh_1^0 \bh_2^0 + 
\bh_1^- \bh_2^+) + {\rm h.c.} \nonumber \\
&\supset& - {1\over 2} 
\left(\begin{array}{cc} 
\tilde h_1^0 & \tilde h_2^0 \end{array}\right)
\left(\begin{array}{cc} 
0 & -\mu \\ -\mu & 0 \end{array}\right)
\left(\begin{array}{c} 
\tilde h_1^0 \\ \tilde h_2^0 \end{array}\right)
- \mu\; \tilde h_1^- \tilde h_2^+ 
+ {\rm h.c.}\;.
\label{eq2}
\eeqa
This Lagrangian defines a degenerate 
spectrum of one charged (Dirac) fermion plus the two
neutral fields
\beqa
\chi_s &=& {i\over \sqrt{2}} (\tilde h_1^0 + \tilde h_2^0) 
\nonumber \\
\chi_a &=& {1\over \sqrt{2}} (\tilde h_1^0 - \tilde h_2^0)\;,
\label{eq3}
\eeqa
all of them with mass $\mu$ (hereafter we assume $\mu>0$). 
The most remarkable feature in this new basis $(\chi_s\; \chi_a)$
is that the gauge couplings with the $Z$ boson become
non-diagonal:
\beqa
{\cal L}_Z &=&-{g\over 2c_W}\; Z_\mu
\left(\begin{array}{cc} 
\overline {\tilde h_1^0} & \overline {\tilde h_2^0} 
\end{array}\right)
\overline \sigma^\mu
\left(\begin{array}{cc} 
1 & 0 \\ 0 & -1 \end{array}\right)
\left(\begin{array}{c} 
\tilde h_1^0 \\ \tilde h_2^0 \end{array}\right) \nonumber \\
&=& -{g\over 2c_W}\; Z_\mu
\left(\begin{array}{cc} 
\overline {\chi_s} & \overline {\chi_a} 
\end{array}\right)
\overline \sigma^\mu
\left(\begin{array}{cc} 
0 & 1 \\ 1 & 0 \end{array}\right)
\left(\begin{array}{c} 
\chi_s \\ \chi_a \end{array}\right)\;.
\label{eq4}
\eeqa

In the MSSM tree-level mixing with the gauginos and loop
(fermion-sfermion and $\gamma/Z$-higgsino) corrections 
\cite{Giudice:1995qk} introduce mass splittings $\Delta_{0,+}$
between the two neutral states and between the
charged and the lightest neutral state  
(which corresponds to 
$\chi_0\approx \chi_a + O(\Delta_0/\mu) \chi_s$ 
if the mixing with the gauginos dominates).
Here, however, sfermions and gauginos 
decouple, and both top-quark loops and the mixing
with the gauginos are negligible. 
Only the splitting 
$\Delta_+ \approx \alpha \mu \log(1+M_Z^2/\mu^2)$
is generated through $\gamma/Z$-higgsino loop corrections. 
Therefore, the typical spectrum in the split-SUSY framework 
with decoupled gauginos consists of two degenerate neutral 
higgsinos of mass $\mu$ plus a charged field around
1 GeV heavier.
 
\section{Higgsinos in the partly-SUSY model}

At the lowest order 
there will be few differences between the scenario just 
described and the partly-SUSY model \cite{Gherghetta:2003wm}. 
The setup is defined in the usual 5D slice of AdS space
of the Randall-Sundrum model \cite{Randall:1999ee}.
It is assumed that the SM fermion and gauge fields live in the 
bulk of the extra dimension, whereas the Higgs fields are attached
to the TeV boundary. SUSY is then broken only in the Planck brane,
and the zero Kaluza-Klein (KK) modes 
of all the SUSY particles in the bulk
(sfermions and gauginos) get large masses.
The fact that SUSY is not broken in the TeV brane would
justify a {\it little} hierarchy between the typical scale there
($L^{-1}\approx 5$ TeV, for example) and the EW scale, since the 
(SUSY-breaking) Higgs 
mass parameters would appear at the loop level suppressed by a 
factor of $(g/4\pi)^2$. The most remarkable feature in this setup 
is that the Higgs sector is (up to low-energy corrections) SUSY 
despite having unsuppressed 
interactions with other sectors of the theory where SUSY may
be broken at the Planck scale. SUSY-breaking contributions 
in the TeV brane involve loops where a bulk particle propagates 
from the TeV brane to the Planck brane and back to the initial
point, and are then red-shifted by the metric.

A first difference with the spectrum in split-SUSY models is 
that here the scalar higgs sector is similar to the one in the 
MSSM, with two neutral and one charged fields in addition 
to the lightest neutral Higgs. 

In the Higgsino sector we assume a $\mu$ term 
(see Eq.~(1)) localized on the TeV brane. In addition, 
the zero modes of sfermions and gauginos are 
very heavy and decouple, which would imply a spectrum 
of charginos and neutralinos that 
coincides (at the lowest order in 1/$L^{-1}$) with the
one described in the split-SUSY case. However, 
the partly-SUSY spectrum also includes the 
KK excitations
of all the fields in the bulk. These fields will be 
localized near the TeV brane, are 
(approximately) SUSY, and have masses of order 
$L^{-1}$.
Their effect can be found using the (SUSY) equations
of motion to integrate them out. In particular,
the KK modes of the vector superfields introduce the 
operator
\beq
{\cal L}_1 = - \int {\rm d}^4\theta\; 
{4.1\over L^{-2}}\; \left( \sum_a g^2 
\left( \bH^\dagger_1 T^a \bH_1 + 
\bH^\dagger_2 T^a \bH_2\right)^2 + {g'^2\over 4} 
\left(- \bH^\dagger_1 \bH_1 + \bH^\dagger_2 \bH_2\right)^2 
\right)\;, 
\label{op1}
\eeq
where the factor of $4.1/L^{-2}$ was calculated in 
\cite{Gherghetta:2003wm} using
the 5D propagator and subtracting out the zero-mode contribution 
(it corresponds to $\sum_n (f_n^2/f_0^2)/M_n^2$, the sum 
over all the excitations of
the inverse mass-squared weighted by the ratio of wave functions 
at the TeV brane). We obtain
\beqa
{\cal L}_1&\supset& 2 \;{4.1\over L^{-2}} \biggl( 
\sum_a g^2 (F_{H_1}^\dagger T^a \tilde H_1\; 
H^\dagger_1 T^a \tilde H_1 + F_{H_2}^\dagger T^a \tilde H_2 \;
H^\dagger_2 T^a \tilde H_2 \nonumber \\
&&\;\;\;\;\;\;\;\;\;\;
+F_{H_1}^\dagger T^a \tilde H_1 \;
H^\dagger_2 T^a \tilde H_2 + F_{H_2}^\dagger T^a \tilde H_2 \;
H^\dagger_1 T^a \tilde H_1) \nonumber \\ 
&&\;\;\;\;\;\;\;\;\;\; + g T^a\rightarrow g' Y \biggr)+{\rm h.c.}\;,
\label{op11}
\eeqa
with $F_{H_1}^\dagger=\mu (h_2^0\; -\! h_2^+)$ and 
$F_{H_2}^\dagger=\mu (h_1^-\; -\!h_1^0)$. 
If 
$\langle h_1^0\rangle = v_1$ and 
$\langle h_2^0\rangle = v_2$ the operator implies the 
higgsino mass terms
\beqa
{\cal L}_1 
&\supset& - {8.2 M_Z^2\over L^{-2}}\;
\biggl[\;{1\over 2}  \;
\left(\begin{array}{cc} 
\tilde h_1^0 & \tilde h_2^0 \end{array}\right)
\left(\begin{array}{cc} 
-\mu\; \sin 2\beta & \mu\; \cos 2\beta \\ 
\mu\; \cos 2\beta& \mu \;\sin 2\beta \end{array}\right)
\left(\begin{array}{c} 
\tilde h_1^0 \\ \tilde h_2^0 \end{array}\right)
\nonumber \\
& &\;\;\;\;\;\;\;\;+ \mu \;\cos^2\theta_W \;\cos 2\beta \; 
\tilde h_1^- \tilde h_2^+
\biggr]+ {\rm h.c.} \;,
\label{op1m}
\eeqa
where $\tan\beta=v_2/v_1$ and $v=\sqrt{v_1^2+v_2^2}=174$ GeV.

In this model the Higgs fields could also interact with
massive chiral superfields localized on the TeV brane. 
These could be the KK modes of bulk fields or purely 4D
fields (in both cases the fields are SUSY and with 
masses of order $L^{-1}$). If trilinear terms 
couple the Higgs doublets with these singlet 
or triplet superfields, integrating them out one obtains the 
effective operator $W= -(\lambda/L^{-1}) (\bH_1 \bH_2)^2$:
\beqa
{\cal L}_2 &=& -\int {\rm d}^2\theta\; {\lambda\over L^{-1}}\; 
\left( \left(\bh_1^0 \bh_2^0\right)^2 -
2 \bh_1^0 \bh_2^0\bh_1^- \bh_2^+ +\left(\bh_1^- \bh_2^+\right)^2
\right) + {\rm h.c.} \nonumber \\
&\supset& - {1\over 2} \;{-2 \lambda\over L^{-1}}\; 
\left(h_2^0 \tilde h_1^0 + h_1^0 \tilde h_2^0 \right)^2
- {2 \lambda\over L^{-1}}\; h_1^0 h_2^0 
\tilde h_1^- \tilde h_2^+ 
+ {\rm h.c.}
\label{op2}
\eeqa
The Higgs vacuum expectation values (VEVs) will then introduce 
an additional mass $-2 \lambda v^2/ L^{-1}$ for the neutral higgsino
$\sin\beta\; \tilde h_1^0 + \cos\beta\; \tilde h_2^0$ and
a mass $\lambda v^2 \sin 2\beta / L^{-1}$ for the chargino.

Several observations are here in order.

{\it (i)} The new mass terms in Eq.~(\ref{op1m}) (from
the integration of gauge excitations)
do not break
the degeneracy between the two neutral higgsinos. The new 
(degenerate) eigenvalue is 
$m_\chi\approx  \mu (1+8.2 \cos 2\beta\; M_Z^2/L^{-2}) $. 
In addition, for $\tan\beta>1$ the mass
contribution to the charged higgsino is negative, making
it lighter than the neutral states. 
To get a working scenario we need
that these effects are compensated by the second operator. 

{\it (ii)} The corrections in Eq.~(\ref{op1m}) are 
of order $v^2 \mu/L^{-2}$, whereas the ones from the integration
of chiral superfields 
(in Eq.~(\ref{op2})) are of order 
$v/L^{-1}$. Therefore, if chiral trilinears and gauge couplings 
are of similar size, for $\mu < L^{-1}$ this second operator
will dominate. A little hierarchy $M_Z\approx \mu \approx 
(g/4\pi) L^{-1}$ is favored in the partly-SUSY model under
discussion \cite{Gherghetta:2003wm}.

{\it (iii)} For a positive value of $\lambda$ (in Eq.~(\ref{op2})),
the second operator defines a spectrum where
the LSP is a neutral state
of mass $m_{\chi}=\mu - \lambda (1-\sin2\beta) v^2/L^{-1}$,
the second neutralino is 
$\Delta_0= 2 v^2 \lambda/ L^{-1}$ heavier, and
the chargino increases its mass to
$m_{\chi^+}=\mu+ \lambda \sin 2\beta\; v^2 / L^{-1}$.

\begin{figure}[!ht]
\centerline{
\includegraphics[width=0.5\linewidth]{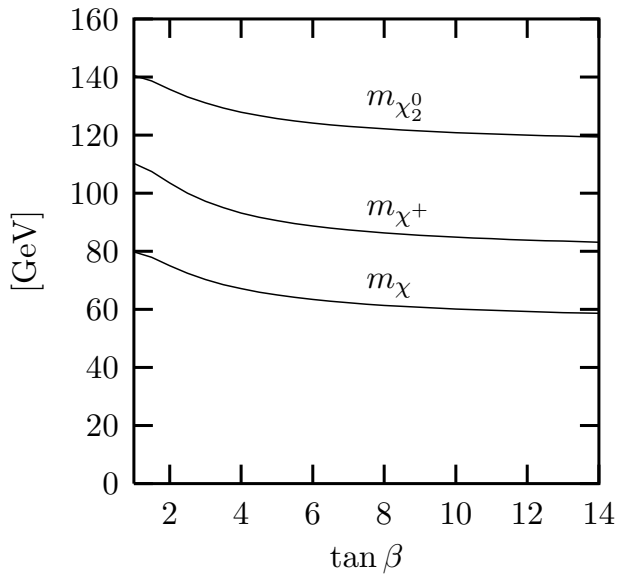}}
\caption{Masses of the LSP ($m_\chi$), the chargino
($m_{\chi^+}$) and the second neutralino ($m_{\chi^0_2}$)
for $\mu=80$ GeV, $L^{-1}=5$ TeV and $\lambda=5$. 
\label{fig1}}
\end{figure} 

In Fig.~1 we plot the three masses for different values of
$\tan\beta$. We take $\mu=80$ GeV, $L^{-1}=5$ TeV and 
$\lambda=5$ (which could be generated by a trilinear coupling 
with a bulk singlet around two times the gauge coupling).
We observe that the corrections are able to keep the LSP lighter
than the $W$ boson while pushing the chargino mass
above the bounds
from LEP. As we see in the next section, this would suffice
to make the neutral higgsino an acceptable dark matter 
candidate.

\section{Dark matter density}

Recent observations \cite{Bennett:2003bz}
indicate that the fraction of critical
energy density of the universe provided by dark matter is
$\Omega_{\chi}=0.113 h^2$. In this section we 
use a modified version of {\it Dark SUSY} \cite{Gondolo:2004sc}
to analize under what 
conditions the higgsinos can account for that number.

The relic density of 
LSP depends crucially on its mass $m_\chi$ and on the rate of the
reactions that change its number \cite{Lee:1977ua}.
If the LSP $\chi$
is significangly lighter than 
the other SUSY particles, then the only relevant reaction is
its anihilation into SM particles. In our framework, 
however, there are other
particles ($\chi^0_2$ and $\chi^+$) with similar mass and
then similar abundances at temperatures below $m_\chi$. These
particles can {\it coanihilate} with $\chi$ 
into SM particles \cite{Griest:1990kh,Mizuta:1992qp}, decreasing 
significantly the freeze-out 
temperature and the relic density of the LSP.

Let us start describing the situation in the split-SUSY case.
If $\mu \lsim M_W$, then the most efficient process 
reducing the LSP abundance is the 
coanihilation with $\chi^0_2$ into quarks and leptons
mediated by a $Z$ boson. For example, taking $\mu=75$ GeV we
obtain $\Omega_{\chi} h^2=0.0005$ (with no significant dependence
on $\tan\beta$). 
Notice that this value
of $\mu$ would be also excluded by collider bounds on the
chargino, which would be just around $1$ GeV heavier. 
If $\mu \gsim M_W$
there is also the anihilation into $W^+W^-$ (with the chargino
in the $t$ channel) and  
into $ZZ$ that push $\Omega_{\chi}$ to 
low values. For example, taking $\mu=95$ GeV we obtain
$\Omega_{\chi} h^2=0.0009$. Therefore, the region with a light higgsino
in the split-SUSY setup can not provide the observed dark 
matter density. For larger values of $\mu$ the 
relic abundance increases, reaching 
$\Omega_{\chi} h^2\approx 0.113$ 
for $\mu=1.1$ TeV (with no significant dependence
on $\tan\beta$).

\begin{figure}[!ht]
\centerline{
\includegraphics[width=0.5\linewidth]{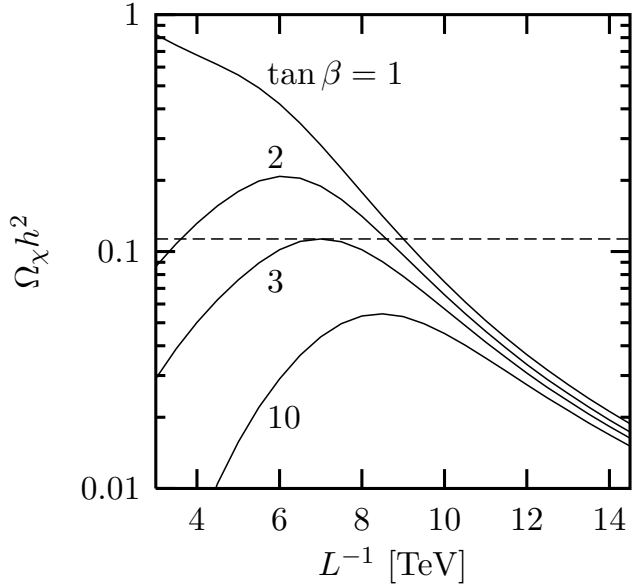}}
\caption{
$\Omega_{\chi} h^2$ for $\mu=75$ GeV, $\lambda=5$ and
different values of $\tan\beta$.
\label{fig2}}
\end{figure}

The situation in the partly-SUSY framework
could be completely different. In particular, 
for $\mu\lsim M_W$ the 
operator in Eq.~(\ref{op2}) can introduce 
splittings that will suppress the relevance of coanihilations
and push the chargino mass above collider bounds.
Let us be more definite. 
If $\tan\beta=1$
the corrections increase the mass of the neutral 
state $\chi^0_2=\chi_s$ (in Eq.~(\ref{eq3}))
to $m_{\chi_2^0}=\mu+\Delta_0$ without changing the 
mass $m_\chi=\mu$ of the LSP $\chi=\chi_a$. 
At temperatures below $\mu$
the coanihilations of $\chi$ and $\chi^0_2$ through a $Z$
will not be relevant
because of the mass splitting (that suppresses the abundance
of $\chi^0_2$), 
whereas the anihilations will
be suppressed because the antisymmetric state does
not couple to the $Z$ boson (the couplings are non-diagonal,
see Eq.~(\ref{eq4})). In Fig.~2 we show
that the corrections are able to increase the relic abundance
up to the observed value. We plot $\Omega_{\chi} h^2$ for
$\mu=75$ GeV and 
different values of $L^{-1}$ and $\tan\beta$. 
For $\tan\beta > 1$ the LSP does not correspond to $\chi_a$,
since the corrections in Eq.~(\ref{op2}) will mix
that state with $\chi_s$. This increases the coupling of
the LSP with the $Z$ boson and its anihilation cross section,
reducing $\Omega_{\chi} h^2$.
Therefore, low values of $\tan\beta$ can accommodate 
larger dark matter densities. 
For a given value of $\tan\beta$, Fig.~2 shows a value of
$L^{-1}$ that optimizes the relic density: larger values 
reduce the mass splittings and increase the relevance of 
coanihilations, whereas lower values of $L^{-1}$ increase
(except for $\tan\beta=1$) the coupling of the LSP with 
the $Z$ boson and then the rate of its anihilations. 

\begin{figure}[!ht]
\centerline{
\includegraphics[width=0.5\linewidth]{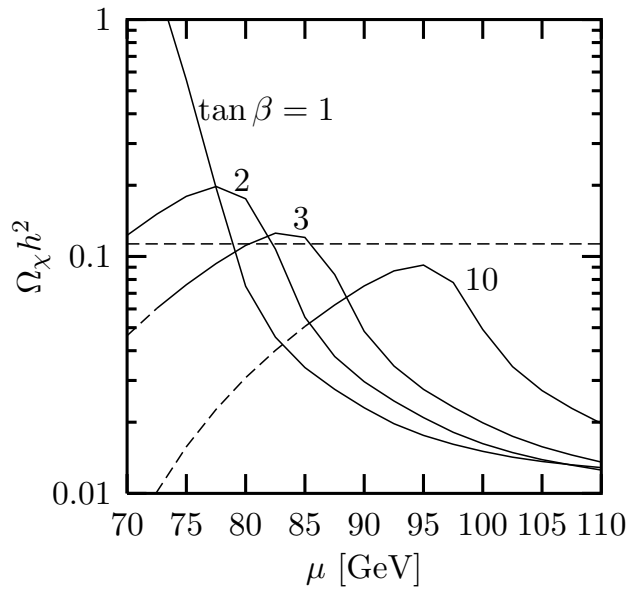}}
\caption{
$\Omega_{\chi} h^2$ for $L^{-1}=5$ TeV, $\lambda=5$ and
different values of $\tan\beta$ and $\mu$. Dashes 
indicate a chargino excluded by collider bounds.
\label{fig3}}
\end{figure}

In Fig.~3 we plot $\Omega_{\chi} h^2$ for different values
of $\mu$ and $\tan\beta$. The maximum value of 
$\Omega_{\chi} h^2$ is always achieved for a LSP mass 
around 75 GeV (larger masses open new anihilation 
channels, and lower masses increase its  
coupling to the $Z$ boson), but due to the mass corrections
this corresponds to different values of $\mu$ 
(depending  on $\tan\beta$). 
We observe in Fig.~3 that for $L^{-1}=5$ TeV and  
$\tan\beta < 3$ the LSP could provide the observed dark
matter of the universe. In general, given $\tan\beta$
and $\lambda$ 
there is a value of $\mu$ and 
$L^{-1}$ that optimizes $\Omega_{\chi} h^2$. For example,
the line corresponding to $\tan\beta=3$ increases up to
$\Omega_{\chi} h^2=0.137$ if $L^{-1}=6.5$ TeV and
$\mu=82$ GeV, whereas for 
$\tan\beta=10$ the maximum value  
$\Omega_{\chi} h^2=0.093$ (within $3\sigma$ deviations of the
experimental value) is achieved for $L^{-1}=5.2$ TeV
and $\mu=95$ GeV. Values of $\tan\beta$ up to 4.7 can
reproduce the central value $\Omega_{\chi} h^2=0.113$.

\section{Summary and discussion}

The LSP has been the favorite WIMP candidate to constitute 
the dark matter of the universe. In particular, the MSSM 
could explain the value $\Omega_{\chi} h^2 = 0.113$ if 
a sneutrino or a neutralino is the LSP. This has  
been an important phenomenological argument for SUSY, in
addition to the basic (and more {\it formal}) motivation of
offering a mechanism to cancel quadratic corrections. 

Recently, however, other scenarios have been 
proposed where SUSY is not the key ingredient to
explain the difference between the EW and the Planck 
scales. We have considered split-SUSY models, where
the higgs mass is the result of an accidental 
cancelation
of much larger contributions, and partly-SUSY models,
where the higgs sector {\it sees} the large SUSY-breaking 
scale red-shifted to the EW scale by the metric.
In both frameworks SUSY would manifest only in
a sector of the theory. Models partially SUSY had not
been studied before because, in general, one expects
that if SUSY is broken in a non-isolated sector 
radiative corrections will extend this breaking to the whole 
theory. This is not the case, however, in the two setups
that we have studied. In split-SUSY 
models the higgsinos could be the only light SUSY fields
if their mass is protected by flavor symmetries or, 
for example, if SUSY breaking
is gauge mediated to sfermions and gauginos while 
the $\mu$ term in generated through gravitional interactions.
In the partly-SUSY case the higgsinos would naturally be
the only light ($\approx (g/4\pi) L^{-1}$) SUSY particles 
if the rest of SM fields live in the bulk 
and SUSY is broken in the Planck brane. 
In both scenarios the
presence of higgsinos could well be the only 
trace of SUSY at energies $\le 1$ TeV.

We have analyzed if these higgsinos could be
the dark matter of the universe. In both scenarios the
degeneracy of the neutral and the charged higgsino states is
the key factor, as coanihilations are then very efective
and imply a too low relic density for $\mu < 1$ TeV. 
In the split-SUSY case the degeneracy is 
only broken (in a $\le 1\%$) by EW loop corrections,
whereas in the partly-SUSY model there are also 
effective operators (suppressed by powers of $1/L^{-1}$) 
that result
after integrating out the KK modes of gauge and chiral fields.

Therefore, although they may look at first 
sight similar, these two models
suggest a very different dark
matter scenario. A 1 TeV LSP, with extra neutral and 
charged fermions around $1$ GeV heavier, together with a
SM content in the scalar sector (no extra higgses), 
would be an indication of split SUSY. In the 
partly-SUSY model it would be more difficult (although
possible) to accommodate 
the 1 TeV LSP, since that would introduce a 
little hierarchy problem. We have shown, however, that 
this framework may imply a LSP 
of mass $m_\chi\approx 75$ GeV, a  
charged higgsino with 
$m_{\chi^+}= m_\chi + \Delta_+\approx 100$ GeV and another
neutral state at 
$m_{\chi^0_2}= \approx m_\chi + 2\Delta_+\approx 125$ GeV. 
Such a higgsino spectrum, with no signs of sfermions or gauginos 
and a scalar higgs sector that 
includes the usual charged and neutral fields of the MSSM
with a low value of $\tan\beta$ ($\lsim 4$), 
would be the clear signature of a partly-SUSY model.

\section*{Acknowledgements}
We would like to thank Mar Bastero, Alex Pomarol and 
Ver\'onica Sanz for useful discussions.
This work has been supported by MCYT (FPA2003-09298-C02-01) and
Junta de Andaluc\'\i a (FQM-101). I.M.
acknowledges a grant from the University of Trieste.

\end{document}